\documentclass[preprint]{aastex}

\usepackage{psfig}

\newcommand{\et}{et~al.\ }
\newcommand{\Ha}{H$\alpha$}

\newcommand{\Msun}{\ensuremath{M_\odot}}

\shorttitle{Extreme X-ray Behavior of NGC 4395}
\shortauthors{MORAN ET AL.}

\received{2004 June 28}
\begin{document}

\title{Extreme X-ray Behavior of the Low-Luminosity Active Nucleus in NGC 4395}

\author{Edward C.\ Moran,\altaffilmark{1}
        Michael Eracleous,\altaffilmark{2}
        Karen M.\ Leighly,\altaffilmark{3}
        George Chartas,\altaffilmark{2}
        Alexei V.\ Filippenko,\altaffilmark{4}
        Luis C.\ Ho,\altaffilmark{5} and 
        Philip R.\ Blanco\altaffilmark{6}}

\altaffiltext{1}{Astronomy Department, Wesleyan University, Middletown, CT
                 06459.}

\altaffiltext{2}{Department of Astronomy and Astrophysics, The Pennsylvania
                 State University, 525 Davey Laboratory, University Park,
                 PA 16802.}

\altaffiltext{3}{Department of Physics and Astronomy, The University of
                 Oklahoma, 440 West Brooks Street, Norman, OK 73019.}

\altaffiltext{4}{Department of Astronomy, University of California,
                 601 Campbell Hall, Berkeley, CA 94720-3411.}

\altaffiltext{5}{The Observatories of the Carnegie Institution of Washington,
                 813 Santa Barbara Street, Pasadena, CA 91101-1292.}

\altaffiltext{6}{Center for Astrophysics and Space Sciences, University of
                 California, San Diego, 9500 Gilman Drive, La Jolla, CA
                 92093-0111.}

\begin{abstract}

We present the results of a 17 ks {\it Chandra\/} ACIS-S observation of the
nearby dwarf spiral galaxy NGC~4395. {\it Chandra\/} affords the first
high-quality, broadband X-ray detection of the active nucleus of this object
that is uncontaminated by nearby sources in the field.  We find the nuclear
X-ray source to be unresolved and confirm the rapid, large-amplitude
variability reported in previous studies.  The light curve appears to show
$\sim 11$ cycles of a quasi-periodic oscillation with a period of $\sim 400$~s.
If associated with an orbital feature near the innermost stable orbit of the
accretion disk, this period would constrain the black-hole mass to be $M<9
\times 10^5\; M_{\odot}$.  The X-ray spectrum indicates absorption by an
ionized medium, and the spectral shape appears to vary over the course of our
observation.  Contrary to prior reports, however, the spectral variations are
uncorrelated with changes in the hard X-ray flux.  It is possible that the
short-term spectral variability we observe results from column density
fluctuations in the ionized absorber.  A power-law fit to the spectrum above
1~keV yields a photon index of $\Gamma \approx 0.6$, much flatter than that
typically observed in the spectra of Seyfert~1 galaxies. We have ruled out
photon pile-up as the cause of the flat spectrum.  Even when complex spectral
features are considered, the photon index is constrained to be $\Gamma < 1.25$
(90\% confidence). Comparing our results with previous determinations of the
photon index ($\Gamma = 1.46$ and 1.72), we conclude that the slope of the
primary continuum varies significantly on time scales of a year or less.  The
extreme flatness and dramatic long-term variability of the X-ray spectrum are
unprecedented among active galactic nuclei.
\end{abstract}


%\newpage
\section{Introduction}

The optical/ultraviolet spectrum of the nucleus of NGC~4395, a nearby dwarf
spiral galaxy, exhibits many of the features that characterize the spectra
of luminous type~1 Seyfert galaxies and quasars (Filippenko \& Sargent 1989;
Filippenko, Ho, \& Sargent 1993).  In addition to a featureless continuum
and strong forbidden emission lines, some of which indicate a very high
degree of ionization, NGC~4395 displays broad permitted emission lines with
velocity widths of several thousand km~s$^{-1}$.  Additional evidence from
across the electromagnetic spectrum (Moran et al.\ 1999; Lira et al.\ 1999;
Kraemer et al.\ 1999; Iwasawa et al.\ 2000; Wrobel, Fassnacht, \& Ho 2001)
supports the hypothesis that NGC~4395 is powered by accretion onto a black
hole, similar to other active galactic nuclei (AGNs).

In a number of respects, however, the AGN in NGC~4395 is a prominent outlier.
Most notably, with an absolute $B$ magnitude of $-11$ and an \Ha\ luminosity
of a few times $10^{38}$ erg~s$^{-1}$, it is the optically least luminous
broad-line AGN currently known.  Based on stellar velocity dispersion
measurements and other arguments (Filippenko \& Ho 2003), the mass of the
black hole in NGC~4395 is constrained to be at most $\sim 6 \times 10^6\, 
\Msun$, and it is likely to be in the  $10^4-10^5\, \Msun$ range --- 
significantly lower than the measured masses of black holes in the nuclei of
other galaxies, both active and quiescent.  Consistent with its low black-hole
mass, the AGN in NGC~4395 is located in a bulgeless dwarf galaxy; in
contrast, other galaxies known to contain nuclear black holes are more massive
and have well-developed bulges (e.g., Kormendy 2001).  NGC~4395 displays
some exceptional emission characteristics as well.  For example, it is one
of the most X-ray--variable active galaxies known (Iwasawa et al.\ 2000),
and its radio--to--X-ray spectral energy distribution differs markedly from
those of other types of AGNs (Moran et al.\ 1999), perhaps indicating that
there is something special about the mode of accretion in NGC~4395.  X-ray
emission is directly linked to the accretion processes in AGNs, and in this
paper, we present the results of a high spatial resolution, broadband
{\it Chandra\/} observation of the central X-ray source in NGC~4395.  The
data provide new insights into the nuclear activity in this unusual object.
We adopt a distance of 4.1 Mpc for NGC~4395 (Thim et al.\ 2004).

\section{The {\it Chandra\/} Observation}

NGC 4395 has been the target of pointed X-ray observations in the 0.1--2.4
keV band with {\sl ROSAT\/} (Moran \et 1999; Lira \et 1999) and twice in the
0.6--10 keV band with {\sl ASCA\/} (Iwasawa \et 2000; Shih, Iwasawa, \& 
Fabian 2003).  However, because
of the modest spectral coverage and collecting area of {\sl ROSAT\/} and
the poor angular resolution of {\sl ASCA}, a complete picture of the
high-energy properties of NGC 4395 is lacking.  {\it Chandra\/} provides
the first broadband, high-resolution X-ray view of this object.

NGC~4395 was observed with {\it Chandra\/} on 20 June 2000 (UT) for 17,187~s
with the ACIS-S instrument.  The S3 chip was the only active CCD during the
exposure; a 1/2
subarray was read out to reduce the possibility of photon pile-up, which
yielded a frame time of 1.54~s.  The nucleus of the galaxy was located at
the instrument aim point.  We reprocessed the data with the CIAO software,
version 2.1.2, in order to remove the 0\farcs25 pixel randomization introduced
during the standard processing of the data.  The screened data consist of
events with grades 0, 2, 3, 4, and 6; no background flares occurred during
the observation.  In the following sections, we describe the X-ray imaging,
variability, and spectroscopy results obtained with {\it Chandra}.

\section{X-ray Imaging}

The {\sl ROSAT\/} PSPC image of NGC~4395 revealed that there are at least
five X-ray sources (labeled A--E by Moran \et 1999) within 2\farcm8 of the
galaxy center, one of which (source~E) is significantly brighter than
the nucleus (source~A) at low X-ray energies.  Thus, in the images obtained
with {\sl ASCA\/} (half-power diameter $\approx 3'$), the nuclear source
was significantly contaminated by emission from the other nearby sources,
particularly below 2 keV (Iwasawa \et 2000).  As Figure~1 illustrates, the
nucleus of NGC~4395 is completely isolated in the $\sim 1''$ resolution
{\it Chandra\/} image.\footnote{As a result of the subarray employed in
our 17~ks exposure, source E, the strongest source of soft X-rays in the
vicinity of the nucleus, was located outside the field of view.  It was,
however, included within the field of a short 1260~s ACIS-S observation
of NGC~4395 (see Ho et al.\ 2001), which we have collected from the {\it
Chandra\/} data archive.  Figure~1 is a superposition of the two datasets;
the true strength of source~E is therefore not represented in the figure.}
The high angular resolution of {\it Chandra\/} also allows us to measure
the position and spatial extent of the nuclear X-ray emission in NGC~4395
with unprecedented accuracy.  The centroid of the nuclear source is at
$\alpha$(2000) = $12^{\rm h} 25^{\rm m} 48.\!\!^{\rm s}$84,
$\delta$(2000) = $33^{\circ} 32' 48.\!\!{''}9$,
just 0\farcs4 from the measured position of the VLBA radio source (Wrobel \et
2001) that is coincident with the optical nucleus.  The X-ray source offset
is well within the $\sim$~1\farcs5 {\it Chandra\/} position uncertainty,
which should lay to rest any lingering doubt about the association between
the central X-ray emission and the nucleus of NGC~4395.  The radial profile
of the nuclear X-ray source indicates that it is unresolved; the fractional
encircled energy within a $2''$ radius is about 97\%.  There is no evidence
for diffuse emission in the vicinity of the nucleus.

A $5''$ radius aperture was used to extract light curves and the spectrum of
the nucleus.  The background was measured in a concentric annulus with inner
and outer radii of $20''$ and $40''$, respectively, which lies between the
nucleus and the nearest off-nuclear source (source B in Fig.~1).  Just 17
counts in the source aperture (0.7\% of the total present) are expected to be
background.  A net total of 2374 source counts was detected for the nucleus
in the 0.3--10.0 keV energy range.  Note that this is a factor of 20 greater
than the number of counts detected in the {\sl ROSAT\/} PSPC exposure of
NGC~4395 (Moran \et 1999), which had almost exactly the same duration.

\section{X-ray Variability}

Previous X-ray observations of NGC~4395 have established the variable nature
of its nucleus.  In the {\sl ROSAT\/} band, the nuclear source varied by a
factor of three over the several days spanning the PSPC observation (Moran et
al.\ 1999; cf.\ Lira et al.\ 1999).  Rapid variability (with doubling times
on the order of 100~s) was witnessed during the {\sl ASCA\/} observation at
energies greater than 2~keV, where the nucleus was more clearly isolated from
other sources in the field (Iwasawa et al.\ 2000). The absence of contamination
in the {\it Chandra\/} data over the 0.3--10 keV range affords a new look at
the variability properties of NGC~4395.  In addition, the light curve derived
from them is continuous, unlike the light curves produced from the
{\sl ROSAT\/} and {\sl ASCA\/} observations.

\subsection{The {\it Chandra\/} Light Curve}

The full-band light curve of the nucleus of NGC 4395 is displayed in Figure~2.
A time bin size of 77~s (= 50 frame times), which yields an average of
$\sim$~10 counts per bin, has been employed.
As the figure indicates, the source was highly variable during the 17~ks
{\it Chandra\/} observation; the count rate fluctuated by an order of
magnitude during the exposure, with dramatic changes (factors of 2--3)
occurring over very short periods.  Several of the strong flares and
dips in the light curve [e.g., the flickering in the $t \approx (0.7 - 1.3)
\times 10^4$~s range] appear to be temporally resolved at this binning.  There
are, however, a few instances (most notably in the $t \approx 7000$--9000~s
range) where the count rate increased and then decreased (or vice versa)
by a factor of $\sim 2$ over the span of a single 77~s bin.  This time
scale constrains the size of the emitting region to be less than 77
light-seconds ($\sim 2 \times 10^{12}$ cm), under the assumption that
the observed fluctuations are associated entirely with changes in the
intensity of the source.

\subsection{Spectral Dependence of the X-ray Variability}

Both {\sl ASCA\/} observations of NGC~4395 seemed to indicate an excess of
soft X-ray emission during periods when the 2--10 keV count rate was high
(Iwasawa et al.\ 2000; Shih et al.\ 2003). We have examined the {\it Chandra\/}
data for a similar trend.  As a first step, we have constructed light curves
in different energy bands: hard (2--10 keV; ``H''), medium (1--2 keV; ``M''),
and soft (0.3--1.2 keV; ``S'').
As shown in the top two panels of Figure~3, the hard-band and full-band light
curves are very similar, both in terms of the features present and the
total intensity.  Thus, most of the detected source counts have energies
above 2~keV.  However, the light curves in the medium and soft energy
bands (third and fourth panels of Fig.~3) provide important additional
information about the variability properties of NGC~4395.  While the
count rates in all three bands are elevated in the $t \approx 7000$--9000~s
range, there are also clear differences between the H, M, and S light
curves. Of particular interest are the dip in the 2--10 keV count rate at
$t \approx 5500$~s, which is absent in the soft and medium energy bands,
and the strong increase in the soft-band count rate at $t \approx$
11,500~s, which is not apparent above 2~keV.  These differences result
in significant spikes in the X-ray ``colors'' of the source (i.e., the
S/H and M/H count-rate ratios), which are shown in the bottom two panels
of Figure~3.  Sharp color variations are evident at other times as well,
but in general they do not appear to correspond directly with changes in
the 2--10 keV count rate.  

We have investigated the issue further, since any dependence of the spectral
properties on count rate would require us to model the source's high-state
and low-state spectra separately.  Unfortunately, given the
degree of structure in the light curve shown in Figure~2, it is difficult to
divide the data into temporally distinct active and quiescent phases, as
Iwasawa et al.\ (2000) did for the {\sl ASCA\/} observation.  However, the
source was clearly more active at certain times during the {\it Chandra\/}
exposure than at others; we therefore extracted an ``active'' spectrum using
data in the $t = 6900$--10,400~s window and a ``quiescent'' spectrum using
data in the $t = 1200$--2400, 5000--6000, and 12,000--16,800~s windows.  These
spectra contain 28\% and 20\% of the total counts, respectively.  As an
alternative approach, we have derived the distribution of count rates in the
light curve, which is displayed in Figure~4.  Based on this histogram, we
have compiled high-state and low-state spectra of the source by collecting
counts in periods represented by the highest 6 bins ($> 0.24$ count s$^{-1}$)
and the lowest 5 bins ($< 0.15$ count s$^{-1}$) separately.  These spectra
contain $\sim 25$\% and 40\% of the detected counts, respectively.  

In Figure~5 we have plotted the ratio of the high-state and low-state
spectra derived using the two methods described above.  Despite the flux
difference of a factor of $\sim 3$ between the two spectra in each case,
spectral variations are only marginally significant.  This is contrary
to the {\sl ASCA\/} results reported by Iwasawa et al.\ (2000) and Shih
et al.\ (2003).  We note that the shape of the spectral ratio shown in
Figure~5$b$, derived using the count-rate histogram in Figure~4, {\it does\/}
bear some resemblance to a similar plot presented by Iwasawa et al.\ (2000).
However, the 1--2 keV excess we observe relative to the 4--10 keV ratio
($\sim 40$\%) is far less than the factor of $\sim 2$ excess indicated
in the {\sl ASCA\/} data.  For the purposes of modeling the ACIS-S spectrum,
we have no evidence that the spectrum changes significantly in direct
response to count-rate fluctuations.  Therefore, the spectral analysis we
present in \S~5 involves the spectrum derived from the full {\it Chandra\/}
data set.

\subsection{Quasi-Periodic Oscillations}

Some of the spikes in NGC~4395's light curve that follow the increase in the
average count rate at $t \approx 7000$~s (see Fig.~2) appear to be evenly
spaced, which has prompted us to search for evidence of periodic variations
in the {\it Chandra\/} data.  As Figure~6$a$ indicates, a periodogram for the
full ACIS-S observation shows a clear peak corresponding to a period of 398~s.
Since the structure associated with this signal seems to be present in only
a portion of the light curve, we have also constructed a periodogram for the
second half of the observation (Fig.~6$b$).  The peak in this periodogram,
now at a period of 396~s, is significantly stronger.  We note that a similar
feature is {\it not\/} present in a periodogram of the first half of the
observation, which suggests that the period is both real (i.e., not an
instrumental effect) and transient.  The 396~s peak in Figure~6$b$ meets all
the qualitative criteria discussed by Halpern, Leighly, \& Marshall (2003)
for a plausible candidate period: (1) it is strong and narrow, (2) it stands
out clearly from surrounding points in the periodogram, and (3) the number
of cycles in the light curve associated with the period --- at least 11 --- is
sufficiently large.

We have adopted the methods of Halpern et al.\ (2003) to determine the
statistical significance of the periodic signal indicated in Figure~6;
full details of the approach are described in that paper.  Essentially, we
simulate light curves using the sampling time, count rate, and variance of
the observed light curve and compare their periodograms to the one obtained
from the data.  Our analysis assumes that the power spectrum of the observed
light curve consists of a power-law component plus a contribution from Poisson
noise.  Simulated light curves for the power-law component were constructed
using the method of Timmer \& K\"onig (1995), in which the Fourier phases and
amplitudes of the power spectrum are randomized prior to generation of the
time series via an inverse fast Fourier transform.  These light curves were
26 times as long as the observation; sections equal in length to the {\it
Chandra\/} observation were extracted at random starting points in each.
We also generated Poisson-noise light curves appropriate for the average
count rate of the observation (10.5 counts in each 77~s bin).  Total
simulated power spectra were then produced by combining individual power
spectra for the two components; the results were rebinned logarithmically
according to the method of Papadakis \& Lawrence (1993).

Two sets of simulations are needed to establish the statistical significance of
features in the periodograms.  The first involves a Monte Carlo determination
of the intrinsic shape of the observed light curve's power spectrum.  For
different combinations of the slope and normalization of the power spectrum's
power-law component, we simulated 1000 light curves and compared their mean
and standard deviation to the observed power spectrum using $\chi^2$.  In
principle, the minimum value of $\chi^2$ should indicate the best estimates
of the slope and normalization.  However, because of the limited duration of
the NGC~4395 observation and the modest count rate of the source, the power
spectrum is dominated by Poisson noise at frequencies greater than
$\sim 10^{-3}$~Hz, where the candidate periodicity is found.  Thus, the
values of the parameters that characterize the power-law component have
little bearing on our results.  Our analysis suggests that values of 1.7 for
the slope and 0.0026 for the normalization are suitable choices, which we
have adopted for the second set of simulations.

Using this prescription for the power spectrum, we simulated 10,000 light
curves and computed their periodograms to establish the 68\%, 95\%, 99\%, and
99.9\% confidence levels plotted in Figure~6.  The confidence levels reflect
the fractions of the simulated light curves that had less power than the
amounts indicated at each frequency.  Following Benlloch et al.\ (2001), we
have computed the local single-trial probability of obtaining the peaks in
Figure~6 by chance: $2.3 \times 10^{-4}$ for the 398~s peak in Figure~6$a$,
and $8.9 \times 10^{-6}$ for the 396~s peak in Figure~6$b$.  The ``global
significance'' (e.g., Benlloch et al.\ 2001), which indicates the probability
that a significant feature could be found at any of the frequencies probed
by the light curve, is estimated by multiplying the local single-trial
probability by the number of independent frequencies in the periodogram.
For our 222-point light curve, there are 111 independent frequencies.  Thus,
the global significance of the period indicated in Figure~6$a$ is 97.6\%.
The global significance of the peak in Figure~6$b$ (55 independent frequencies)
is 99.95\%.  The actual global significance may be somewhat lower, for two
reasons.  First, in this analysis, we have oversampled the periodograms in
order to search for periodicity at frequencies that are not equal to the
Fourier frequencies.  The peak in the periodogram for the second half of
the observation, it turns out, falls almost exactly midway between two Fourier
frequencies.  Analyzing the light curve only at the Fourier frequencies, we
obtain a global significance of 97.6\% for this peak.  Second, it has been
suggested that uncertainty in the continuum model for the power spectrum
(in this case, the Poisson noise level) can also decrease the significance
of peaks in the power (Vaughan 2004).  In combination, the inclusion of this
uncertainty and removal of the effects of oversampling could reduce the
significance of the peak to the 95\% confidence level (S.~Vaughan 2004,
private communication).  Even at the lower significance levels, however,
this remains one the best candidates to date for quasi-periodic oscillations
in an AGN X-ray light curve.

\section{X-ray Spectral Analysis}

X-ray spectroscopy provides vital clues about the emission and absorption
components present in the nucleus of NGC~4395, as well as its intrinsic
luminosity.  Interesting spectral properties of this source have already
been revealed in previous X-ray observations.  For example, even though the
signal-to-noise ratio ($S/N$) of the {\sl ROSAT\/} PSPC spectrum was poor,
we were unable to
obtain a good fit using a power-law model with absorption by a Galactic
column of neutral material (Moran \et 1999), which in the {\sl ROSAT\/}
band describes the X-ray spectra of most broad-line AGNs (e.g., Walter
\& Fink 1993).  The absence of strong spectral variability corresponding
to the factor-of-three change in the source flux suggested to us that a
single emission component dominates below 2 keV; we further speculated
that an ionized (or ``warm'') absorber (Halpern 1984) might be responsible
for the spectral complexity.  The subsequent {\sl ASCA\/} observation of
NGC~4395 (Iwasawa \et 2000) confirmed the presence of an ionized absorber
and indicated that it may have multiple components.

For model fitting, the {\it Chandra\/} spectrum of NGC~4395 was grouped into
bins containing a minimum of 25 counts.  Because of uncertainties in the
ACIS-S response at the lowest and highest energies, we have ignored channels
below 0.5 keV and above 9 keV.  The XSPEC software (Arnaud 1996) was used for
spectral modeling.

\subsection{Power-Law Fit}

A power-law model with absorption by neutral material provides a poor fit
to the {\it Chandra\/} spectrum over the entire 0.5--9 keV range.  It does,
however, provide an excellent fit to the data above $\sim 1.2$ keV.  The
best-fit photon index and absorption column density (with 90\% confidence
ranges for one parameter of interest) are
$\Gamma =0.61^{\scriptscriptstyle +0.16}\!\!\!\!\!\!\!\!\!\!
             _{\scriptscriptstyle -0.15}$
and $N_{\rm H} = (1.2 \pm 0.2) \times 10^{22}$ cm$^{-2}$.  The fit, which
indicates an apparent excess of flux below 1~keV, is displayed in Figure~7.
Modeling the spectrum above 2~keV we obtain the same best-fit value for the
photon index.  An unresolved Fe~K$\alpha$ emission line at 6.4~keV is very
marginally detected with an equivalent width of $99 \pm 95$ eV (90\%
confidence).  The improvement to the fit obtained with the inclusion of this
component ($\Delta \chi^2 = 2.98$ for one additional free parameter) is
significant at the 95\% confidence level.

The hard X-ray spectrum of NGC~4395 observed with {\it Chandra\/} is
substantially flatter than the spectra of more luminous broad-line AGNs,
which typically have slopes in the $\Gamma$ = 1.7--1.9 range (e.g., Nandra
\& Pounds 1994).  In addition, the spectral index we have measured is much
lower than the value of $\Gamma = 1.72$ reported by Iwasawa et al.\ (2000)
from their analysis of the first {\sl ASCA\/} observation of NGC~4395.  We
return to this point in \S~6.  Interestingly, the 2--10 keV flux measured
with {\it Chandra\/} ($3.7 \times 10^{-12}$ erg cm$^{-2}$ s$^{-1}$) is
similar to that obtained with {\sl ASCA\/} ($4.5 \times 10^{-12}$
erg cm$^{-2}$ s$^{-1}$).

\subsection{Pile-Up Analysis}

Although the average ACIS-S count rate is just $\sim 0.14$~count~s$^{-1}$,
it is high enough at times to raise concerns about photon pile-up, i.e., the
occurrence of two or more events in a pixel within one frame time.  The effect
of pile-up is to replace events at lower energies with fewer numbers of events
at higher energies, which decreases the apparent count rate of the source and
artificially flattens its spectrum. It can also alter the photon event grades.
A thorough analysis of the pile-up issue is needed to confirm that the
unusually flat spectrum of NGC~4395 we have observed --- potentially one of
its most important properties --- is of a physical, not instrumental, origin.

The first indication that photon pile-up is not significant in our
{\it Chandra\/} observation of NGC~4395 is the fact that the strong edge in
the mirror response near 2.1~keV is visible in the spectrum (see Fig.~7);
this feature is usually smeared out when pile-up is severe.  As a preliminary
test for spectral distortions associated with pile-up, we have analyzed a
spectrum of NGC~4395 that excludes events from the innermost portion of the
source's spatial profile, which pile-up would affect the most.  We extracted
counts in an annulus centered on the source peak with an inner radius of
0.5 pixels (0\farcs25) and an outer radius of 10 pixels ($5''$); the resulting
spectrum consists of 73\% of the total events.  The effective-area file used
when modeling the annular spectrum was corrected for the energy dependence
of the {\it Chandra\/} point-spread function.  Fitting the annular spectrum
with an absorbed power-law model above $\sim 1$~keV, we obtain a spectral
index of
$\Gamma = 0.55^{\scriptscriptstyle +0.19}\!\!\!\!\!\!\!\!\!\!
              _{\scriptscriptstyle -0.26}$.
Thus, the slope of the spectrum measured when the events most likely to be
piled up are excluded is just as flat as that obtained using all the events,
which demonstrates that time-averaged ACIS-S spectrum of NGC~4395 has not
been distorted by pile-up.

To investigate the time dependence of pile-up effects during the observation,
we have performed simulations with the LYNX software (Chartas et al.\ 2000;
Eracleous et al.\ 2002).  LYNX is a forward-fitting tool that simulates the
distribution of counts per energy channel for an assumed input spectrum; using
the timing information for a data set, it iteratively refines the intrinsic
values of the model parameters, accounting for pile-up, until the simulated
count distribution matches the observed one.  LYNX assesses pile-up in two
ways: (1) it provides an estimate of the pile-up fraction, i.e., the fraction
of ``true'' events in the observed band that are lost because of pile-up, and
(2) it provides estimates of the {\it intrinsic\/} values of the spectral
parameters for a given model, which can be compared to those obtained directly
via standard model fitting with XSPEC.  The simulations can be performed for
an entire observation or a portion of it.

The input spectrum employed for our analysis is a double power-law model,
which provides a good fit to the ACIS-S spectrum over the full 0.5--9~keV
range while introducing the fewest possible number of free parameters.
In this model, a single component dominates above 1~keV.  (A more physical
description of the {\it Chandra\/} spectrum is presented in the next section.)
Our LYNX simulations confirm that pile-up effects are minimal: the pile-up
fraction is just 3\% during periods of relative quiescence [e.g., the $t =
(1.2-1.7) \times 10^4$~s range], and during the most active phase (i.e., the
$t$ = 7000--9000~s range), the pile-up fraction is still only $\sim$~10\%.
Pile-up fractions of 10\% or less have negligible effects on the observed
spectrum and are generally considered to be acceptable.  In addition, the
simulations indicate an {\it intrinsic\/} spectral index of
$\Gamma =0.61^{\scriptscriptstyle +0.25}\!\!\!\!\!\!\!\!\!\!
             _{\scriptscriptstyle -0.26}$
for the dominant power law when the source was in an active phase
($t$ = 6000--12,000~s), and
$\Gamma =0.52^{\scriptscriptstyle +0.19}\!\!\!\!\!\!\!\!\!\!
             _{\scriptscriptstyle -0.30}$
during a more quiescent period ($t$ = 0--6000~s).  Both values are very
similar to the observed spectral index of the source (\S~5.1).  Changes
in the parameters associated with the second power law are insignificant
as well.  It is therefore safe to conclude that photon pile-up has not
seriously impacted the spectrum of NGC~4395 at any point during the
{\it Chandra\/} observation, contrary to speculations by Shih et al.\ (2003).

\subsection{Complex Spectral Features}

As Figure~3 indicates, the variability of NGC~4395 at high and low photon
energies, though not identical, is similar, implying that a single emission
component dominates the spectrum of the object over the entire 0.3--10~keV
range.  However, an absorbed power-law model only provides a good fit above
1.2~keV.  To investigate the spectral complexity at low energies, we refitted
the {\it Chandra\/} spectrum above 2.2 keV with a power-law model, assuming
only the Galactic neutral hydrogen column density.  The residuals associated
with this fit, displayed in Figure~8, exhibit several strong dips in the soft
X-ray band, suggesting that an ionized absorber (e.g., Halpern 1984; Reynolds
1997) is present in NGC~4395.  Indeed, with the addition of a warm absorber
to the model (i.e., the ``absori'' component in XSPEC), an excellent fit to
the {\it Chandra\/} spectrum over the full energy range is obtained.  The
best-fit photon index, $\Gamma = 0.63$, is almost identical to that found
above in \S~5.1 with simple power-law fits to the hard-band spectrum.
Fixing the photon index at that value, the parameters associated with the
warm absorber (with uncertainties corresponding to $\Delta \chi^2 = 2.71$)
are as follows: a column density of
$N_{\rm w} = (1.8 \pm 0.3) \times 10^{22}$ cm$^{-2}$
and an ionization parameter of
$\xi = 130^{\scriptscriptstyle +40}\!\!\!\!\!\!\!
          _{\scriptscriptstyle -30}$ erg cm s$^{-1}$.
The high ionization (and reduced opacity) of the absorber is needed for the
model to account for the apparent soft excess indicated in Figure~7.

In their analysis of the $\sim 300$~ks {\sl ASCA\/} observation of NGC~4395,
Shih et al.\ (2003) fitted the high-energy portion of the spectrum with a
complex model consisting of a power-law continuum, Compton reflection, and
a relativistically broadened iron line.  The line and reflection components
contribute flux above $\sim 6$~keV, which leads to a steeper photon index
than that obtained with a simple power-law fit.  Given the extremely flat
spectral index we find for NGC~4395, we should consider the effects of
these additional spectral components --- they could impact the warm absorber
parameters we measure as well.  Unfortunately, because our {\it Chandra\/}
spectrum has only about a tenth as many 2--10~keV counts as the long {\sl
ASCA\/} SIS spectrum, we are unable to obtain meaningful constraints on the
numerous free parameters associated with the additional components.  We can,
however, adopt the reflection and line parameters obtained by Shih et al.\
(2003) at face value as a means of assessing the range of possible values
for the power-law and warm-absorber parameters during the {\it Chandra\/}
observation.  We first fitted the ACIS-S spectrum above 2~keV in exactly
the same manner as described in \S~4.1 of Shih et al.\ (2003); the model
consists of power-law and reflection components (the ``pexrav'' model in
XSPEC), a relativistically broadened iron line (the ``diskline'' model),
and neutral absorption.  Fixing all the parameters (except the power-law
normalization) at the values derived or adopted by Shih et al.\ --- this
includes the photon index of $\Gamma = 1.46$ they measured --- we obtain
a very poor fit to the {\it Chandra\/} spectrum ($\chi^2 = 114$ for 60
degrees of freedom). With the photon index and column density as free
parameters, an excellent fit is possible.  The best fit ($\chi^2$/d.o.f.\
= 45/58) yields the following values:
$\Gamma = 0.93 \pm 0.32$
and
$N_{\rm H} = (1.7 \pm 1.0) \times 10^{22}$ cm$^{-2}$.
Next, we set the neutral absorption to the Galactic value and the photon index
to $\Gamma = 0.93$; the addition of a warm absorber once more affords a good
fit over the full 0.5--9~keV range, yielding the following parameters for the
ionized medium:
$N_{\rm w} = (2.1 \pm 0.2) \times 10^{22}$ cm$^{-2}$
and
$\xi = 50^{\scriptscriptstyle +15}\!\!\!\!\!\!\!
             _{\scriptscriptstyle -10}$ erg cm s$^{-1}$.
The ionization parameter is somewhat lower in this model compared to that
obtained when the hard-band spectrum is represented as a simple power law
with $\Gamma = 0.6$.  This occurs because there is a greater amount of
intrinsic soft X-ray flux in the the $\Gamma = 0.93$ model; a lower
ionization (greater opacity) is needed for a good fit to the spectrum at
low energies.

\section{Implications for the Nature of the Nucleus of NGC 4395}

\subsection{Broadband Luminosity of the Source}

Using the high angular resolution of {\it Chandra}, we have established that
the central X-ray source in NGC~4395 is coincident with the galaxy's optical
and radio nucleus.  The X-ray emission is spatially unresolved and highly
variable, leaving no doubt that it is associated with the dwarf AGN in the
galaxy.  Provided that a single emission component dominates in the
{\it Chandra\/}
band, the power-law fit to the ACIS-S spectrum above 1.2 keV suggests an
absorption-corrected, time-averaged 2--10 keV luminosity of $8.0 \times
10^{39}$ erg~s$^{-1}$ for an adopted distance of 4.1 Mpc.  This is within
25\% of the value obtained with {\sl ASCA\/} (adjusted for this distance)
by Iwasawa \et (2000).  In the 0.5--10 keV band, $L_{\rm X} = 9.0 \times
10^{39}$ erg~s$^{-1}$.  Combining the broadband X-ray luminosity with the
radio--to--UV luminosity computed by Moran \et (1999, after revision for
the 4.1 Mpc distance), we obtain a bolometric luminosity of $L_{\rm bol} =
5.3 \times 10^{40}$ erg~s$^{-1}$.  This luminosity corresponds to an
Eddington-limit mass of $\sim 400$ $\Msun$ for the accreting black hole.
As reported by Filippenko \& Ho (2003), the conservative upper limit for
the black-hole mass is $6 \times 10^6$ $\Msun$; the actual mass, they
argue, is likely to be between $\sim 1 \times 10^4$ $\Msun$ and $\sim 1
\times 10^5$ $\Msun$.  Using this range of values, the bolometric luminosity
suggests an Eddington ratio of $L_{\rm bol}/L_{\rm Edd} = 0.004$--0.04.

\subsection{X-ray Fibrillation in the Heart of NGC 4395}

The nucleus of NGC 4395 is an exceptionally variable X-ray source.  As pointed
out by Iwasawa \et (2000), such extreme variability is quite uncharacteristic
of low-luminosity AGNs (LLAGNs).  We find the ``excess variance'' (Nandra
\et 1997; Turner \et 1999) of the source in the 2--10 keV band is $0.35 \pm
0.03$ (using 128~s bins), almost twice that estimated by Iwasawa \et using
{\sl ASCA\/} data and far greater than the excess variances measured for
other LLAGNs (e.g., Ptak \et 1998).  However, many of the LLAGNs examined
to date are LINERs, which are likely to be accreting in an advection-dominated
mode (e.g., Lasota et al.\ 1996; Ho 1999) --- in other words, they are
supermassive black holes ($M > 10^6$ $\Msun$) with low accretion rates
($\dot{M}/\dot{M}_{\rm Edd} < 10^{-3}$).  In contrast, NGC 4395 appears
to possess a less massive black hole that is accreting at a higher relative
rate.

Amidst the rapid, large-amplitude variability, there is evidence of
quasi-periodic oscillations in the light curve of NGC~4395.  As discussed in
\S~4.3, a strong signal at a period of $\sim 400$~s is seen in periodograms
of the {\it Chandra\/} data.  The period is transient --- a dozen or so cycles
over a duration of $\sim 5$~ks were observed --- but our simulations indicate
that its statistical significance is high.  Among the few previous reports
of periodic variability in AGNs (Halpern et al.\ 2003, and references therein),
this is perhaps the most convincing detection to date.  If real, the periodic
behavior could reveal important clues about the nature of NGC~4395.  For
instance, the other AGNs suspected of having periodic variations,
RX~J0437.4--4711 and Ton~S180 (Halpern et al.\ 2003), have much higher 2--10
keV luminosities ($> 10^{43}$ erg~s$^{-1}$) and far longer periods ($\sim
10^5$~s) than NGC~4395, suggesting a relationship between period and black-hole
mass.  Such a relationship would be expected if the periodic signal is
associated with the dynamical time scale of the accretion disk.  For example,
the period $P$ (in seconds) of an object in Keplerian orbit ``$x$''
gravitational radii ($r_{\rm g} = GM/c^2$) from the black hole implies a
black-hole mass of $M = 3.4 \times 10^4\, P\, x^{-3/2}\, M_{\odot}$.  If the
source of the periodicity in NGC~4395 is located near the innermost stable
orbit of the accretion disk (i.e., $x \approx 6$), a mass of $M \approx 9
\times 10^5\, M_{\odot}$ is implied.  An orbital feature occurring farther
out in the disk (at $x > 6$) would require a lower mass for a fixed period
of 400~s, so this might be taken as a firm upper limit on the mass of the
black hole in NGC~4395.

\subsection{Dramatic Long-Term Spectral Variability}

Of great importance is the possibility that the shape of the broadband
continuum in NGC~4395 has undergone dramatic changes during the three
years that span the {\sl ASCA\/} and {\it Chandra\/} observations.  Taken
at face value, simple absorbed power-law fits to the data above 2~keV
signify that such spectral variability {\it has\/} occurred: 
$\Gamma = 1.72^{\scriptscriptstyle +0.24}\!\!\!\!\!\!\!\!\!\!
             _{\scriptscriptstyle -0.27}$
during the first {\sl ASCA\/} observation (1998 May; Iwasawa et al.\ 2000),
$\Gamma = 0.61^{\scriptscriptstyle +0.25}\!\!\!\!\!\!\!\!\!\!
             _{\scriptscriptstyle -0.26}$
during our {\it Chandra\/} observation (2000 June), and $\Gamma \approx 1$
during the second {\sl ASCA\/} observation (2001 May; Shih et al.\ 2003). We
have also examined the spectrum of the nuclear source in NGC~4395 during the
short 1260~s {\it Chandra\/} ACIS-S exposure obtained by Ho et al.\ (2001),
which was carried out about two months before our 17~ks observation.  The count
rate of the source during the short exposure was just 0.095~count~s$^{-1}$,
so severe pile-up is not a concern.  Despite
the limited $S/N$ ratio of the snapshot spectrum, an absorbed power-law model
provides a good fit above $\sim 1.2$ keV; we find that $\Gamma < 1.0$ (90\%
confidence), consistent with the results of the 17~ks exposure.  The neutral
column densities associated with these simple power-law fits are all in the
narrow range of $(1.2-2.3) \times 10^{22}$~cm$^{-2}$.  Thus, it would appear
that the spectral variations observed above 2~keV are due to large changes in
the effective photon index, not the effective absorption.  The apparent range
of the photon index variations ($\Gamma \approx 0.6-1.7$) is unprecedented
among AGNs.

The possibility of complex spectral features in NGC~4395 at both high and low
X-ray energies complicates the question of whether the intrinsic slope of the
primary continuum varies, or whether the slope is constant and modified to a
varying degree by extrinsic absorption.  From their analysis of the second
{\sl ASCA\/} observation, Shih et al.\ (2003) have concluded that the intrinsic
power law in NGC~4395 has a relatively steep (and presumably constant) slope of
$\Gamma = 1.46^{\scriptscriptstyle +0.02}\!\!\!\!\!\!\!\!\!\!
              _{\scriptscriptstyle -0.06}$
and that its flatter appearance at times is due to changes in the ionization
of the warm absorber.  However, not only is the photon index obtained from the
first {\sl ASCA\/} observation almost excluded by the second measurement, it
was obtained with a simple power-law model --- not one consisting of a power
law, reflection, and a broad iron line, such as that used to derive the value
of 1.46.  As discussed above, inclusion of line and reflection components leads
to a steeper power-law index.  Accounting for these in the first observation
would, if anything, yield an intrinsic photon index that is even steeper than
$\Gamma = 1.72$.  Thus, it seems unlikely that a change in the ionization of
the warm absorber alone could reconcile the different results obtained with
{\sl ASCA}.

We have performed two tests to explore whether the flat spectrum observed
with {\it Chandra\/} could be consistent with the $\Gamma = 1.46$ power law
measured by Shih et al.\ (2003).  The first is based on analysis presented
in \S~5.3, where we fitted the {\it Chandra\/} spectrum above 2~keV using
the reflection and iron line parameters derived by Shih et al.  If the
contributions from these components were the same during the {\it Chandra\/}
observation, we should obtain a steep photon index similar (within the errors)
to the {\sl ASCA\/} result.  As discussed above, the exact model parameters
given by Shih et al.\ provide a poor fit to the {\it Chandra\/} spectrum.
However, allowing the power-law index and neutral column density to vary,
we obtain an excellent fit; the best-fit column density is approximately
the same as that found by Shih et al., but we measure a much flatter spectral
index of $\Gamma = 0.93 \pm 0.32$.  The differences between the {\it Chandra\/}
and {\sl ASCA\/} results are illustrated in Figure~9, which strongly suggests
that a change in the intrinsic photon index has occurred.

As a second test, we have fitted the {\it Chandra\/} spectrum over the full
0.5--9~keV range with the complete model of Shih et al.\ (2003).  In addition
to Compton reflection and broad iron-line emission, this model includes two
components of ionized absorption.  The presence of a second absorber, although
not required by the {\it Chandra} data, cannot be ruled out either, and it
could allow a steep photon index to be accommodated.  Once again, because of
the limited $S/N$ ratio of our data, we are unable to constrain
the numerous free parameters associated with this model, so we have adopted
the best-fit parameters obtained by Shih et al., including the photon index
of $\Gamma = 1.46$.  The only free parameters in the fit were the column
densities and ionization parameters of the absorbers and the normalization
of the continuum.  Interestingly, this model yields a formally acceptable
fit to the full {\it Chandra\/} spectrum. The best-fit column densities we
obtain ($N_{\rm w1} = 1.3 \times 10^{22}$~cm$^{-2}$ and $N_{\rm w2} = 
7.3 \times 10^{22}$~cm$^{-2}$) are similar to those found by Shih et al.\
when they modeled the time-averaged {\sl ASCA\/} spectrum; the best-fit
ionization parameters, on the other hand, are much lower ($\xi_1 = 0.9$ 
erg cm s$^{-1}$ and $\xi_2 = 79$ erg cm s$^{-1}$).  However, we obtain a
much better fit if we allow the photon index to be a free parameter
($\chi^2$/d.o.f. = 64.2/77).  An F-test indicates that the improvement to the
fit ($\Delta \chi^2 = 11.4$ for one additional free parameter) is significant
at a confidence level far greater than 99\%.  The best-fit warm absorber
parameters do not change much ($N_{\rm w1} = 1.2 \times 10^{22}$~cm$^{-2}$,
$N_{\rm w2} = 4.6 \times 10^{22}$~cm$^{-2}$, $\xi_1 = 3.4$ erg cm s$^{-1}$,
$\xi_2 = 116$ erg cm s$^{-1}$).  The photon index of
$\Gamma = 1.10^{\scriptscriptstyle +0.15}\!\!\!\!\!\!\!\!\!\!
              _{\scriptscriptstyle -0.25}$
we obtain clearly excludes the Shih et al.\ value of 1.46.  Once more,
it appears that the intrinsic spectral index of NGC~4395 during the
{\it Chandra\/} observation was less than that observed with {\sl ASCA}.

Without question, the overall shape of the X-ray spectrum of NGC~4395 varies
on time scales of a year or less.  The preponderance of evidence indicates that
changes in the intrinsic slope of the primary continuum are at least partly
responsible.  If real, such dramatic photon index variations --- previously
unobserved in AGNs --- are important for understanding the nature of this
object.  Current theory is challenged to explain the large spectral index
changes and extremely flat power-law slope implied for NGC~4395
(e.g., see Haardt, Maraschi, \& Ghisellini 1997).  We note that the
photon index of $\Gamma \approx 1$ we obtain when reflection and iron
emission are considered is consistent with the minimum slope expected in
the standard picture for hard X-ray production in AGNs, whereby low-energy
photons from an accretion disk are inverse-Compton scattered by a corona
of hot ($kT \approx 1\,\rm MeV$) thermal electrons with low optical depth
($\tau \approx 0.01$; Haardt \& Maraschi 1993).  The time scales for all
relevant processes in the nucleus of NGC~4395 are 2--3 orders of magnitude
shorter than in luminous AGNs.  Thus, the effects of fluctuations of
certain physical conditions (e.g., the spectrum of the seed photons or
temperature of the corona) might be more readily observed in this object.
At any rate, the X-ray spectral variability of NGC~4395 should be monitored;
it may provide new insight into the accretion processes not only in this
object, but perhaps generally in AGNs.

\subsection{Ionized Absorption and the Short-Term Spectral Variability}

The other significant difference between our results and those obtained with
{\sl ASCA\/} concerns the variability of NGC~4395 in the 1--2 keV band and
its dependence on the source count rate at higher energies.  A strong excess
1--2 keV flux associated with increases in the 2--10 keV count rate was seen
in the {\sl ASCA\/} data (Iwasawa et al.\ 2000; Shih et al.\ 2003).  
To account for this behavior, Iwasawa et al.\ (2000) suggested that two
ionized absorbers are present, one that is constant and one that is variable.
In this picture, the ionization parameter of the latter component increases
when the source is in a high state, resulting in a lower opacity in the
1--2 keV range and excess observed flux in this band.  The {\it Chandra\/}
light curves in Figure~3 and the spectral ratio plots shown in Figure~5 do
not indicate this sort of behavior.  In general, excess soft X-ray emission
does not accompany an increase in the hard-band count rate.  In fact, the
largest spike in the (1--2~keV)/(2--10~keV) count-rate ratio (bottom panel
of Fig.~3) coincides with a sharp drop in the hard X-ray flux.

While these results are inconsistent with the details of the multi-zone warm
absorber model discussed by Iwasawa et al.\ (2000), it is important to note
that a constant, single-zone absorber model --- which provides an excellent
fit to the {\it Chandra\/} spectrum of NGC~4395 --- also cannot account for
the short-term spectral variability indicated in Figure~3.  In that
circumstance, no fluctuations in the (1--2~keV)/(2--10~keV) count-rate ratio
should be observed.  We therefore consider the possibility that the
ionized absorbing medium in NGC~4395 is non-uniform or clumpy: The spatial
scale of the clumpiness and the transverse velocity of such a medium could,
in the right combination, give rise to rapid column density variations along
the line of sight, and thus, to changes in the observed soft X-ray flux that
are partially independent of the continuum variability above 2 keV. Soft-band
and hard-band light curves that are similar in some respects, but not in
others, would be a natural consequence in this scenario.  A similar idea
was considered by Reynolds \& Fabian (1995) as a source of spectral
variability in AGNs.

As a preliminary test, we have used the PIMMS software to compute the ACIS-S
count rates that would be observed in various bands if the $\Gamma = 0.6$
power law were absorbed only by a Galactic column of neutral material.
Relative to the 2--10 keV band, we would expect 0.65 and 0.73 times as many
counts in the 0.3--1.2 keV and 1--2 keV bands, respectively.  If variations
in column density associated with a clumpy absorbing medium are responsible
for differences in the soft-band and hard-band light curves we observe, the
X-ray colors of the source defined in Figure~3 should not exceed these amounts.
Referring to the figure, we see that the maximum value of Color~1 (the
[0.3--1.2 keV]/[2--10 keV] ratio) is $\sim 0.4$ and the maximum value of
Color~2 (the [1--2 keV]/[2--10 keV] ratio) is $\sim 0.7$, consistent with
our hypothesis.

To investigate the role of column density variations further, we have used
the Color~1 light curve to predict the values of Color~2 for the entire
{\it Chandra\/} observation.  This was done by adjusting the column density
of the ionized medium in the best-fit $\Gamma = 0.6$ warm absorber model
(see \S~5.3)
until the observed values of Color~1 were obtained.  The same model, with
the modified column density, was then used to derive Color~2.  The implied
range of column densities is $\sim 1 \times 10^{22}$ to $\sim 4 \times 
10^{22}$ cm$^{-2}$.  As Figure~10 shows, the agreement between the observed
and predicted values of Color~2 is quite good.  Although the model we have
employed may not be the best physical description of the source's spectrum,
our results suggest that column density fluctuations may indeed contribute
to the short-term spectral variability indicated in Figure~3.

A few simple calculations can be made to assess the plausibility of the
non-uniform column density scenario.  Following Reynolds \& Fabian (1995),
the time scale of the associated spectral variability would be $t \approx 
\Delta R/v$, where $\Delta R$ is the characteristic size of the
non-uniformities and $v$ is the transverse velocity of the medium.  From
the bottom two panels of Figure~3, we can see that $t \approx 500$~s or
less.  Placing the warm absorber within or just outside the broad-line
region (BLR) --- an ad hoc assumption, but one consistent with the findings
of other investigations (e.g., Reynolds \& Fabian 1995) --- we can estimate
the transverse velocity.  The broad H$\alpha$ line in NGC~4395 has a
full-width at zero-intensity of $\sim 7000$~km~s$^{-1}$ (Filippenko \&
Sargent 1989).  This implies an upper limit of $v = 3500$~km~s$^{-1}$ if
the BLR is made up of spherically distributed clouds with random orbits.
If instead the BLR gas has a flattened distribution that is inclined to
our line of sight, 3500~km~s$^{-1}$ might be a typical velocity; we adopt
it as such.  Thus, $\Delta R \approx 2 \times 10^{11}$~cm is obtained.
The column density of the medium is $N
\approx n \Delta R \approx 2 \times 10^{22}$~cm$^{-2}$, which suggests a
number density $n \approx 10^{11}$~cm$^{-3}$.  If $v$ is a virial velocity,
the location of the warm absorber relative to the central black hole is
$R = GM/v^2$, where $M$ is the mass of the black hole.
This places the ionized medium at $R \approx 10^{13}$~cm if  $M =
10^4\, M_{\odot}$, and at $R \approx 10^{14}$~cm if $M = 10^5\, 
M_{\odot}$.  In either case, $R >> \Delta R$, as it must be.  Assuming the
ionizing luminosity to be $L \approx 10^{40}$~erg~s$^{-1}$, we can now
estimate the value of the ionization parameter $\xi = L/nR^2$: $\xi \approx
1000$ erg cm s$^{-1}$ if $M = 10^4\, M_{\odot}$, and $\xi \approx 10$
erg cm s$^{-1}$ if $M
= 10^5\, M_{\odot}$.  Despite the rough numbers used here, we obtain a
range of values of the ionization parameter that, for the probable range of
the black-hole mass, brackets the various measured values of $\xi$ (\S~5.3;
Iwasawa et al.\ 2000; Shih et al.\ 2003).  Thus, an
association between the short-term spectral variability observed with
{\it Chandra\/} and fluctuations in the column density of the warm absorber
cannot be ruled out.

A question naturally arises: could some of the absorption be due to the BLR
gas itself?  The calculations above suggest that a connection between the
absorber and the BLR gas, at least the more highly ionized portion of it,
is plausible.  Absorbers with physical conditions that are consistent with
those expected for gas in the BLR have been identified in some other more
luminous Seyfert galaxies.  Examples include Mrk~509 (Kriss et al.\ 2000),
NGC~5548 (Kaastra et al.\ 2002), NGC~3783 (Blustin et al.\ 2002), and
IRAS~13349+2438 (Sako et al.\ 2001).  In the case of NGC~4395, photoionization
modeling by Kraemer et al.\ (1999) has indicated that the covering factor of
the BLR gas is very high.  Thus, it is entirely possible that BLR gas lies
along our line of sight to the continuum source, where the X-ray absorber is
located.

\section{Summary and Conclusions}

The nucleus of NGC~4395 is unique among type~1 AGNs.  In addition to having
one of the lowest observed luminosities and least massive black holes, it
displays a number of extreme properties in the X-ray band as well.  As our
{\it Chandra\/} observation confirms, NGC~4395 has a spectacular X-ray light
curve that exhibits large-amplitude variations on extremely short time scales.
This object is one of the most X-ray--variable AGNs known.  In addition, the
{\it Chandra\/} light curve shows possible evidence of quasi-periodic
oscillations.  About a dozen cycles with a period of $\sim 400$~s were
observed.  The statistical significance of the periodicity is high, making
this one of the most convincing cases to date of periodic X-ray variability
in an AGN.  Under the assumption that the feature responsible for the periodic
signal is located near the innermost stable orbit of the accretion disk, the
400~s period implies an upper limit of $9 \times 10^5\; M_{\odot}$ for the
mass of the black hole in NGC~4395, which is consistent with previous
estimates of the black-hole mass.

One of the most intriguing aspects of the nuclear X-ray source in NGC~4395 is
the slope of its hard X-ray spectrum.  The {\it Chandra\/} data indicate an
effective power-law photon index of $\Gamma = 0.6$, which is significantly
flatter than the X-ray spectral slopes associated with more luminous type~1
AGNs ($\langle\Gamma\rangle \approx 1.8$).  Even with the additional spectral
components considered by Shih et al.\ (2003), we find $\Gamma < 1.25$ (90\%
confidence).  In addition, the collective results of recent broadband X-ray
observations of NGC~4395 indicate that the intrinsic slope of its continuum
emission undergoes dramatic changes ($\Gamma \le 1.25$ to $\Gamma \ge 1.7$)
on time scales of a year or less.  Spectral variations of this magnitude
have never been observed in an AGN.

Ionized gas is apparently responsible for the bulk of the X-ray absorption
in NGC~4395.  Although short-term spectral variability is indicated over the
course of the {\it Chandra\/} exposure, we do not observe a strong excess
1--2 keV flux corresponding to increases in the 2--10 keV count rate, which
was the impetus for the complex multi-zone warm absorber model employed by
Iwasawa et al.\ (2000) and Shih et al.\ (2003).  We find that column density
fluctuations associated with a non-uniform absorbing medium are able to
account for the short-term spectral variability we observe.  This is not
to say that the changes in ionization reported in the {\sl ASCA\/} studies
did not occur --- the ionization and column density may both vary in a
real warm absorber.  Some other property of NGC~4395 --- perhaps the slope
of the continuum --- may cause one effect to dominate over the other in
terms of the short-term spectral variability of the source.  We note that
the flux of photons above 1~keV incident on the absorbing medium during the
first {\sl ASCA\/} observation (when $\Gamma \ge 1.72$) was as much as 2.5
times greater than it was during the {\it Chandra\/} exposure.  Such changes
in the ionizing photon flux could significantly affect the physical state
of the absorber.

Obviously, X-ray monitoring of NGC~4395 with high-spatial resolution
instrumentation is crucial for a clarification of the physical processes
occurring in its nuclear region.  Ultimately, a firm understanding of the
unusual behavior in this object may provide valuable insight into the
physics of AGNs.

\acknowledgments

We are grateful to Simon Vaughan for sharing the results of his timing
analysis with us.  We would also like to thank Andrzej Zdziarski for
helpful discussions about X-ray continuum models.  The work of M.E.\ was
supported by NASA grant NAG~5-10817.  A.V.F.\ aknowledges support through
{\it Chandra\/} grant GO0-1170A and NASA grant NAG~5-3556.

\clearpage
\begin{figure}
\begin{center}
\centerline{\psfig{figure=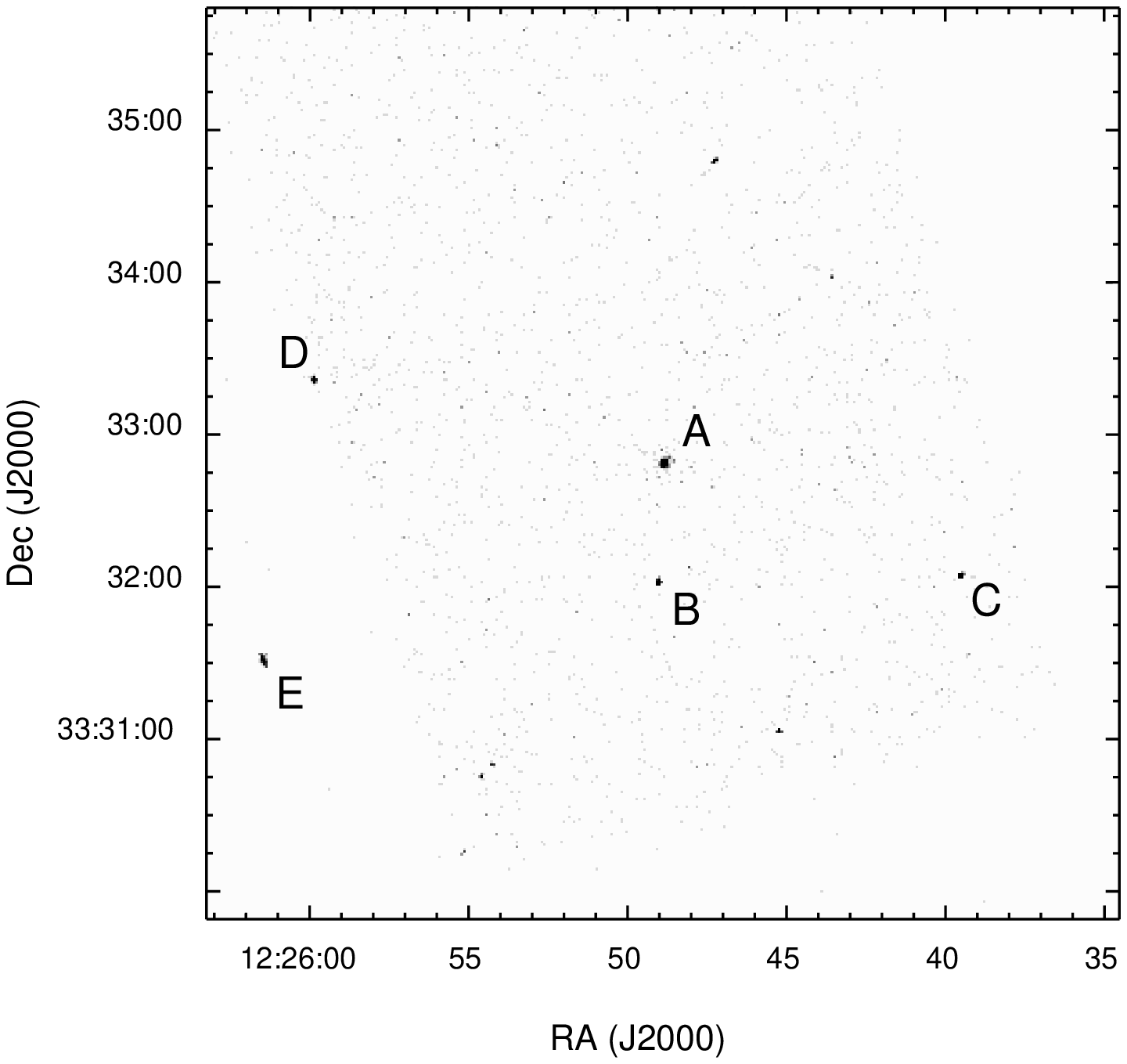,width=8truein,angle=0}}
\caption{{\it Chandra\/} ACIS-S image of NGC 4395 in the 0.3--10 keV
band.  Sources A--E were previously identified in a {\sl ROSAT\/} PSPC
observation (Moran et al.\ 1999); several other sources are visible as well.
Source~A is coincident with the nucleus of the galaxy.  It is clear from
the image that the nucleus is completely isolated from the other sources
in the field, which was not the case in previous X-ray observations of
this object.}
\end{center}
\end{figure}

\clearpage
\begin{figure}
\begin{center}
\centerline{\psfig{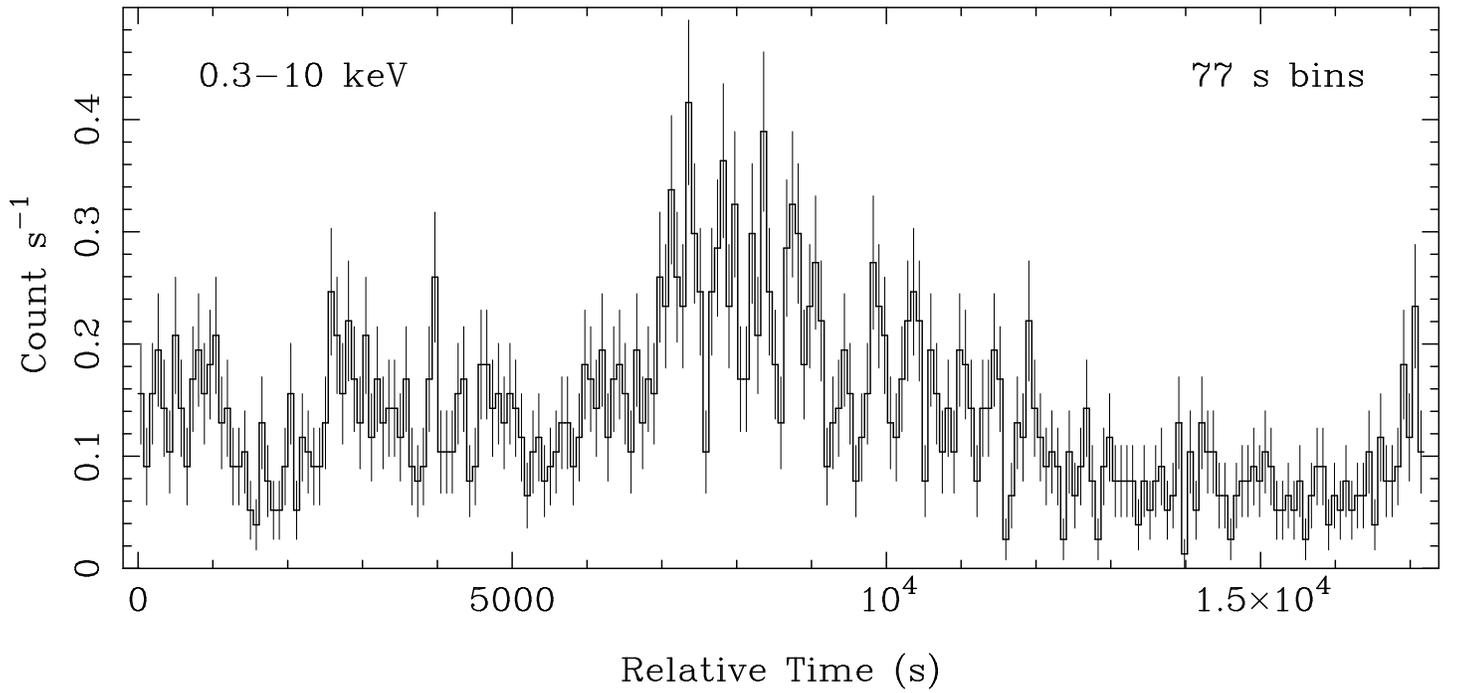}}
\caption{The 0.3--10 keV light curve of NGC~4395.  Each 77~s bin 
corresponds to 50 frames.  The source clearly exhibits large-amplitude
variability on very short time scales.  The error bars shown here and
in subsequent figures represent 1$\sigma$ uncertainties.}
\end{center}
\end{figure}

\clearpage
\begin{figure}
\begin{center}
\centerline{\psfig{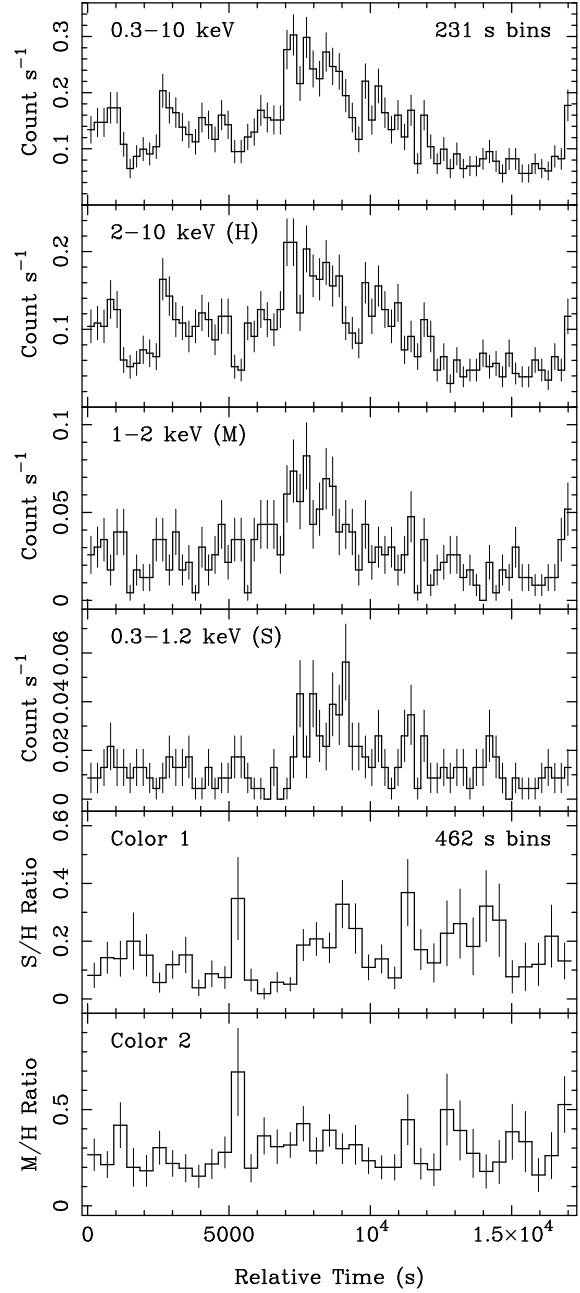}}
\caption{Light curves of NGC 4395 in the broad, hard (H), medium (M), and
soft (S) energy bands (top four panels).  The 231~s bins correspond
to 150 frames.  The bottom two panels indicate the S/H count ratio
(Color 1) and the M/H count ratio (Color 2) for the source in 300-frame
bins.  Note that there is little apparent correspondence between either
color and the 2--10 keV count rate.}
\end{center}
\end{figure}

\clearpage
\begin{figure}
\begin{center}
\centerline{\psfig{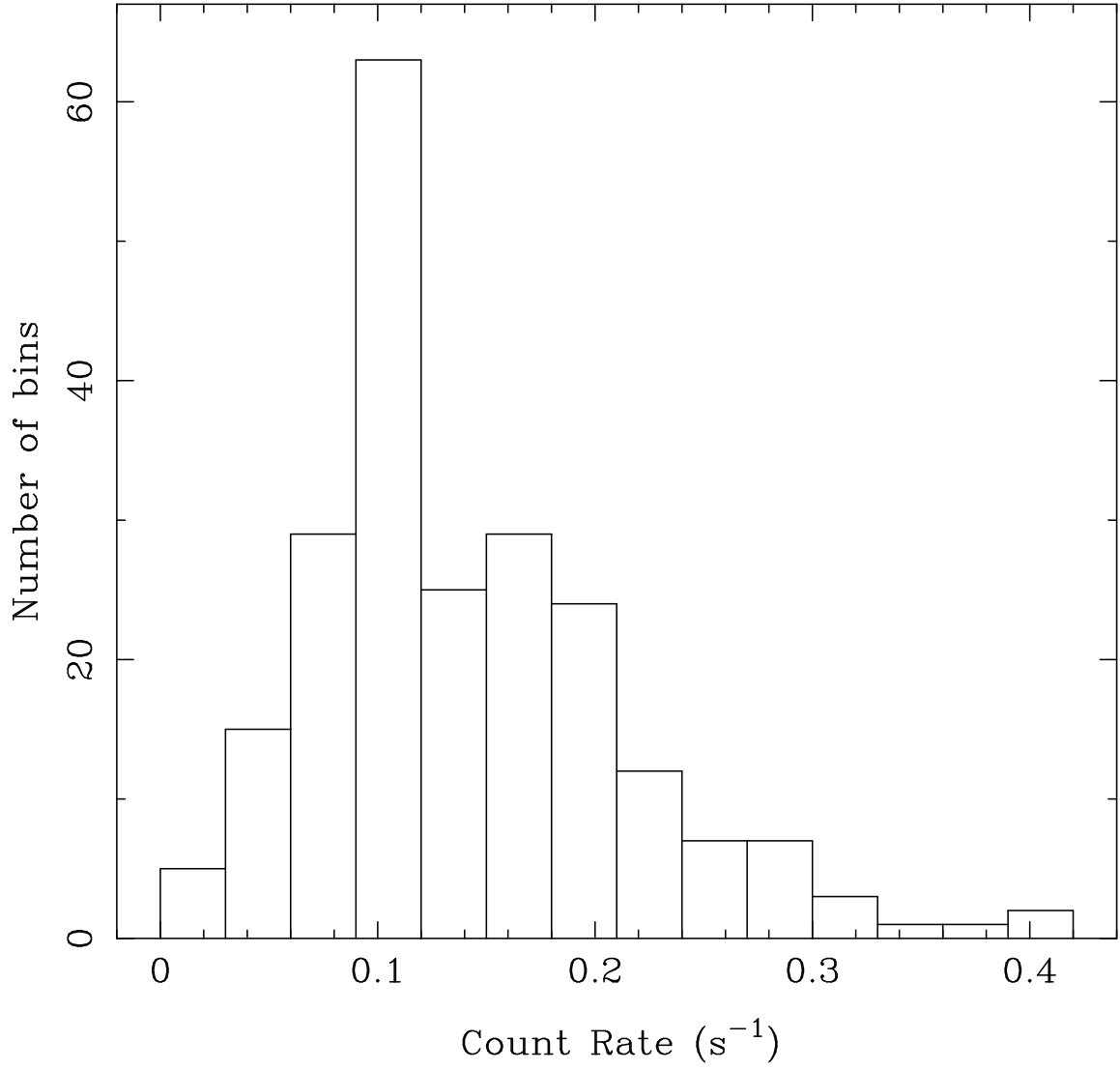}}
\caption{Distribution of count rates in the 50-frame light curve shown
in Fig.~2.}
\end{center}
\end{figure}

\clearpage
\begin{figure}
\begin{center}
\centerline{\psfig{figure=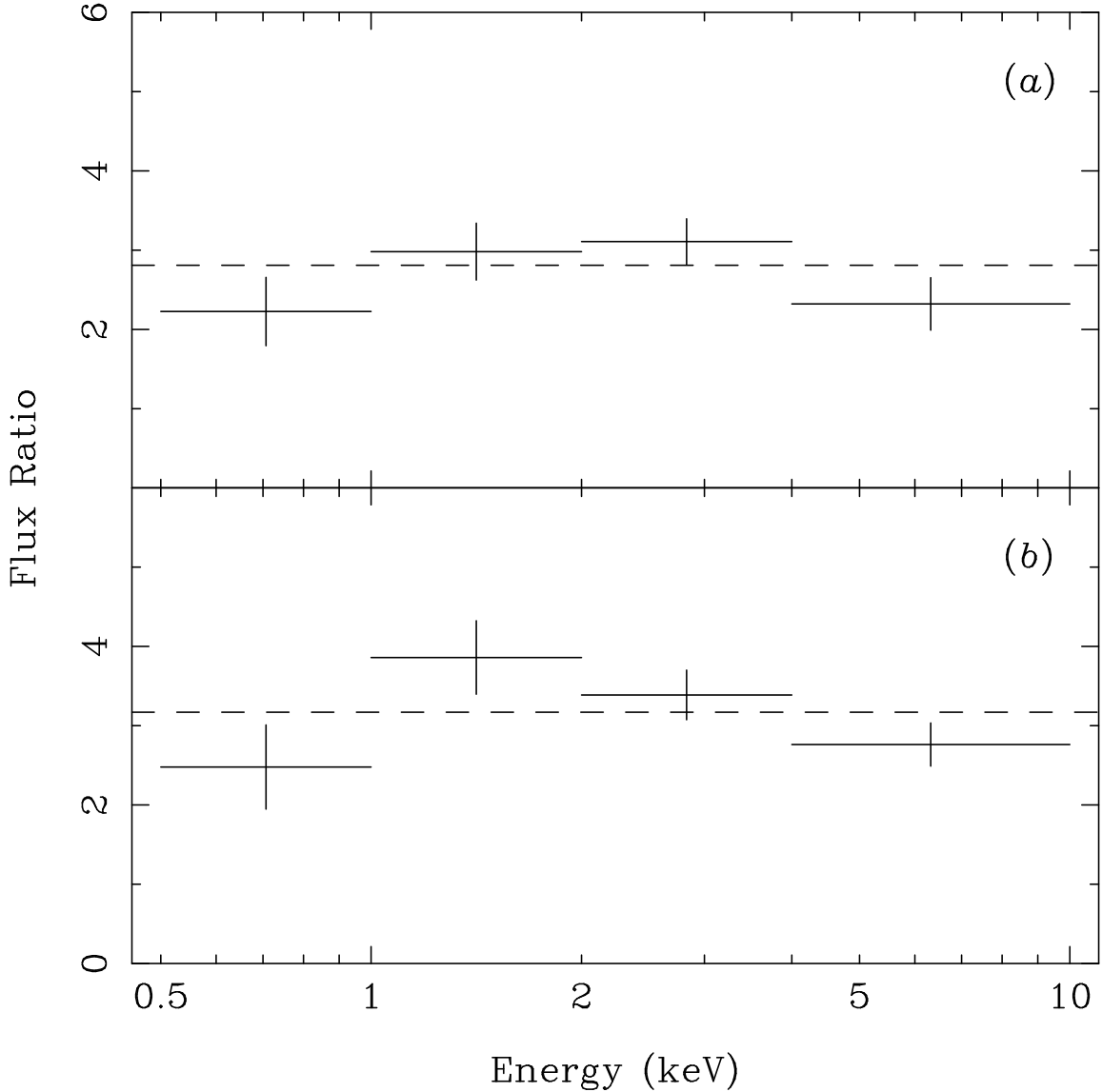,width=6truein,angle=0}}
\caption{Ratio of spectra of NGC 4395 in ``active'' and ``quiescent''
states, which have been defined in two different ways (see \S~5.3 of the text
for details).  In ($a$), the high-state and low-state spectra are associated
with broad periods in the light curve during which the activity of the source
was visibly different.  For ($b$), the high-state and low-state spectra were
derived on the basis of the count-rate histogram shown in Fig.~4.  The dashed
lines denote the mean ratio in the two cases.  Neither plot indicates strong
spectral variability corresponding to the factor-of-three count rate changes.}
\end{center}
\end{figure}

\clearpage
\begin{figure}
\begin{center}
\centerline{\psfig{figure=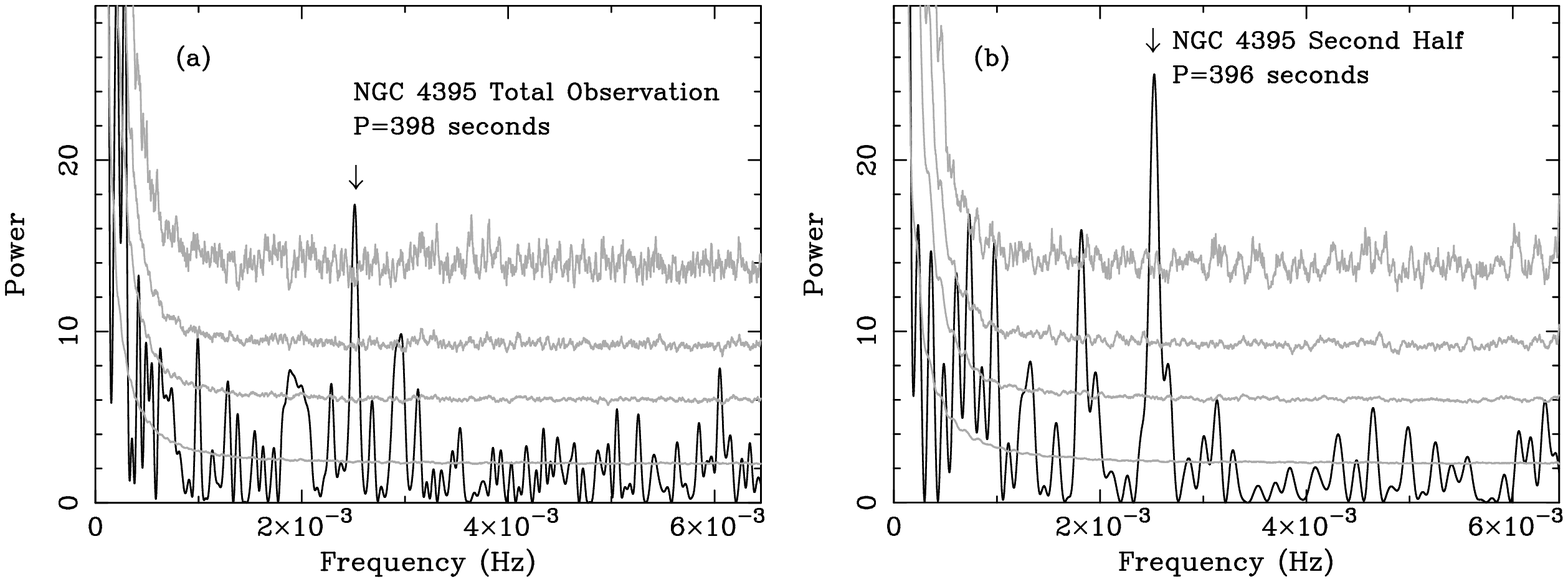,width=6.5truein,angle=0}}
\caption{($a$) Periodogram for the entire 17~ks {\it Chandra\/} observation
of NGC~4395.  The possible period at $P = 398$~s is marked. ($b$) The
significance of the periodic signal is enhanced in the periodogram of
the second half of the observation.  The peak is now at $P = 396$~s.
Overlaid on both plots are the confidence levels determined from 10,000
simulated periodograms, as described in the text.  From bottom to top,
the curves represent 68\%, 95\%, 99\%, and 99.9\% confidence that a signal
exceeding the indicated amount of power is significant.}
\end{center}
\end{figure}

\clearpage
\begin{figure}
\begin{center}
\centerline{\psfig{figure=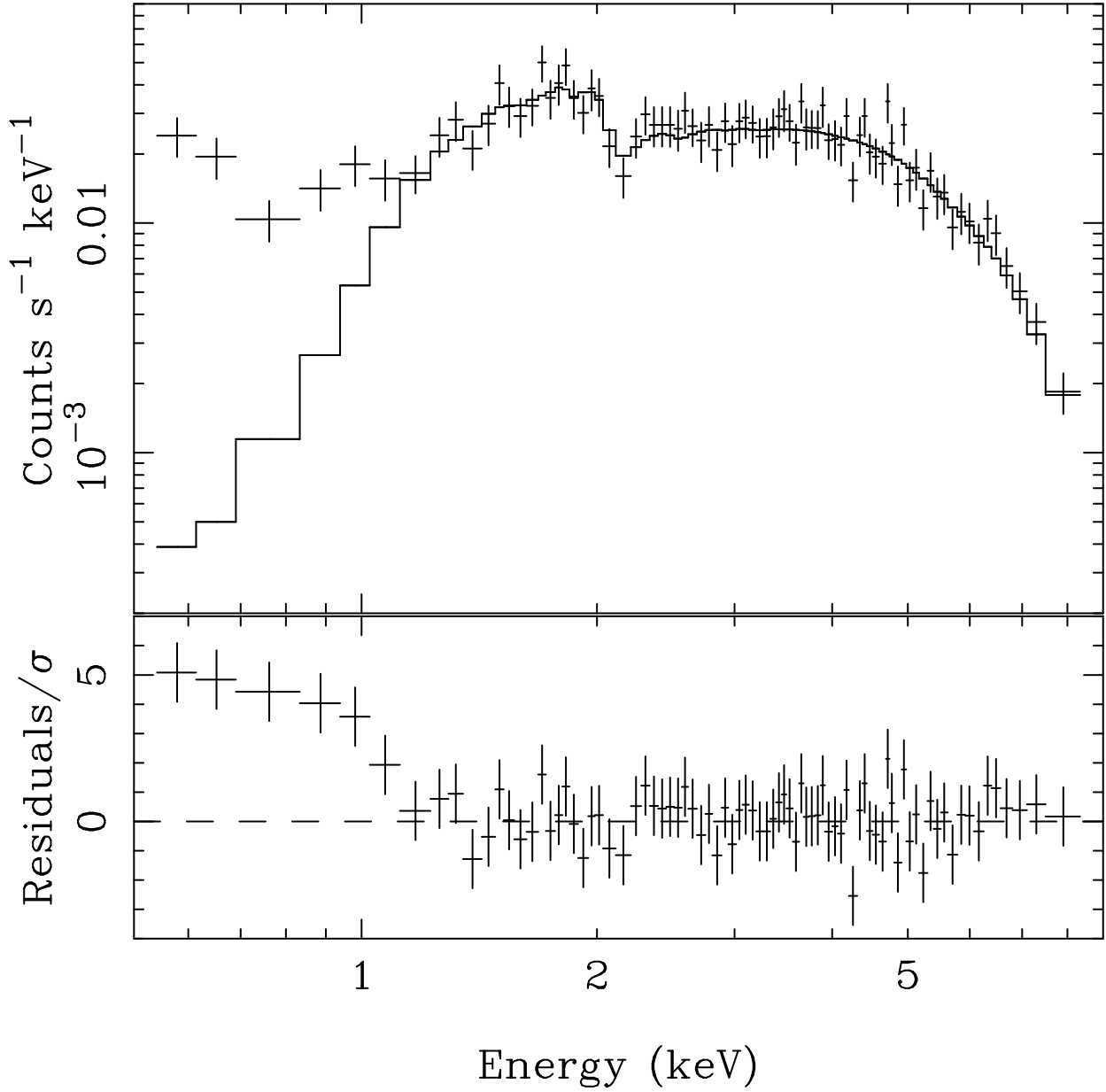,width=6.5truein,angle=0}}
\caption{({\it top panel}) Observed ACIS-S spectrum of NGC 4395, fitted
with a simple absorbed power-law model above 1.2 keV ($\Gamma = 0.61, N_{\rm H}
= 1.2 \times 10^{22}$ cm$^{-2}$).  The fit residuals (normalized by the
1$\sigma$ errors) are shown in the lower panel.  Note the strong excess
flux at the lowest energies.}
\end{center}
\end{figure}

\clearpage
\begin{figure}
\begin{center}
\centerline{\psfig{figure=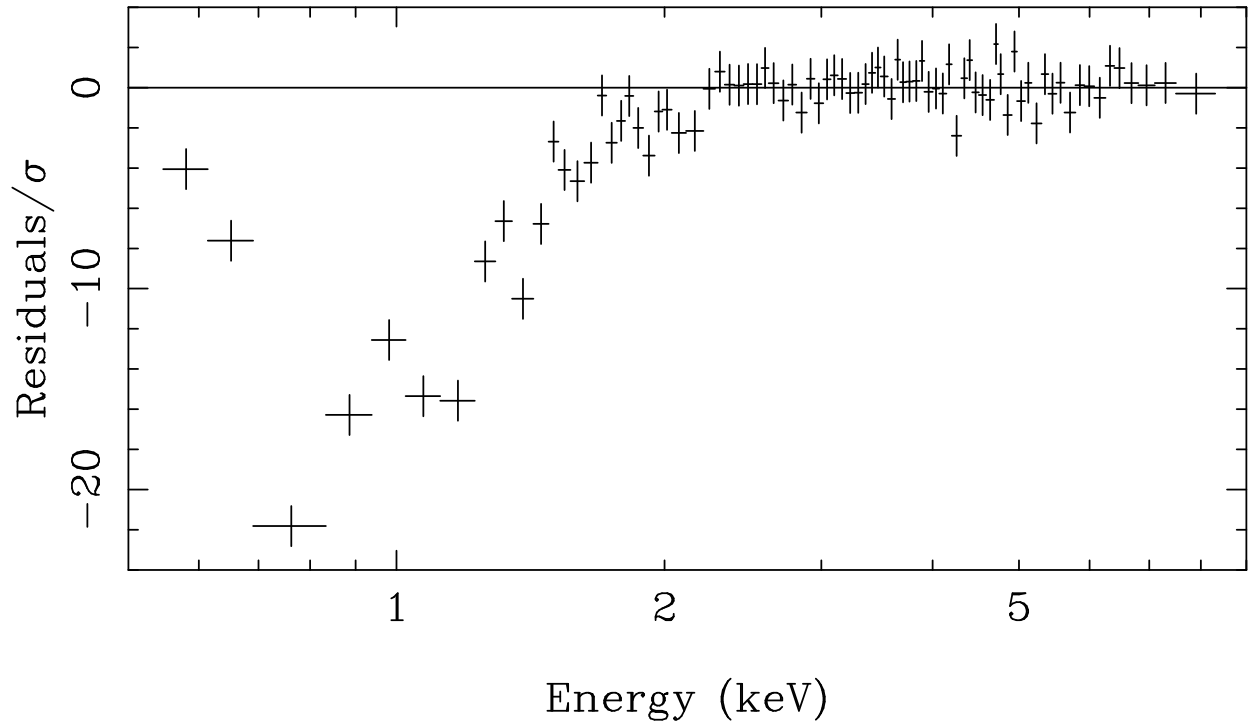,width=6.5truein,angle=0}}
\caption{Normalized residuals from a power-law fit to the ACIS-S spectrum
of NGC 4395 above 2.2 keV.  The absorption has been fixed at the Galactic
value.  The deep troughs seen below 2 keV suggest additional absorption
by an ionized medium.}
\end{center}
\end{figure}

\clearpage
\begin{figure}
\begin{center}
\centerline{\plotone{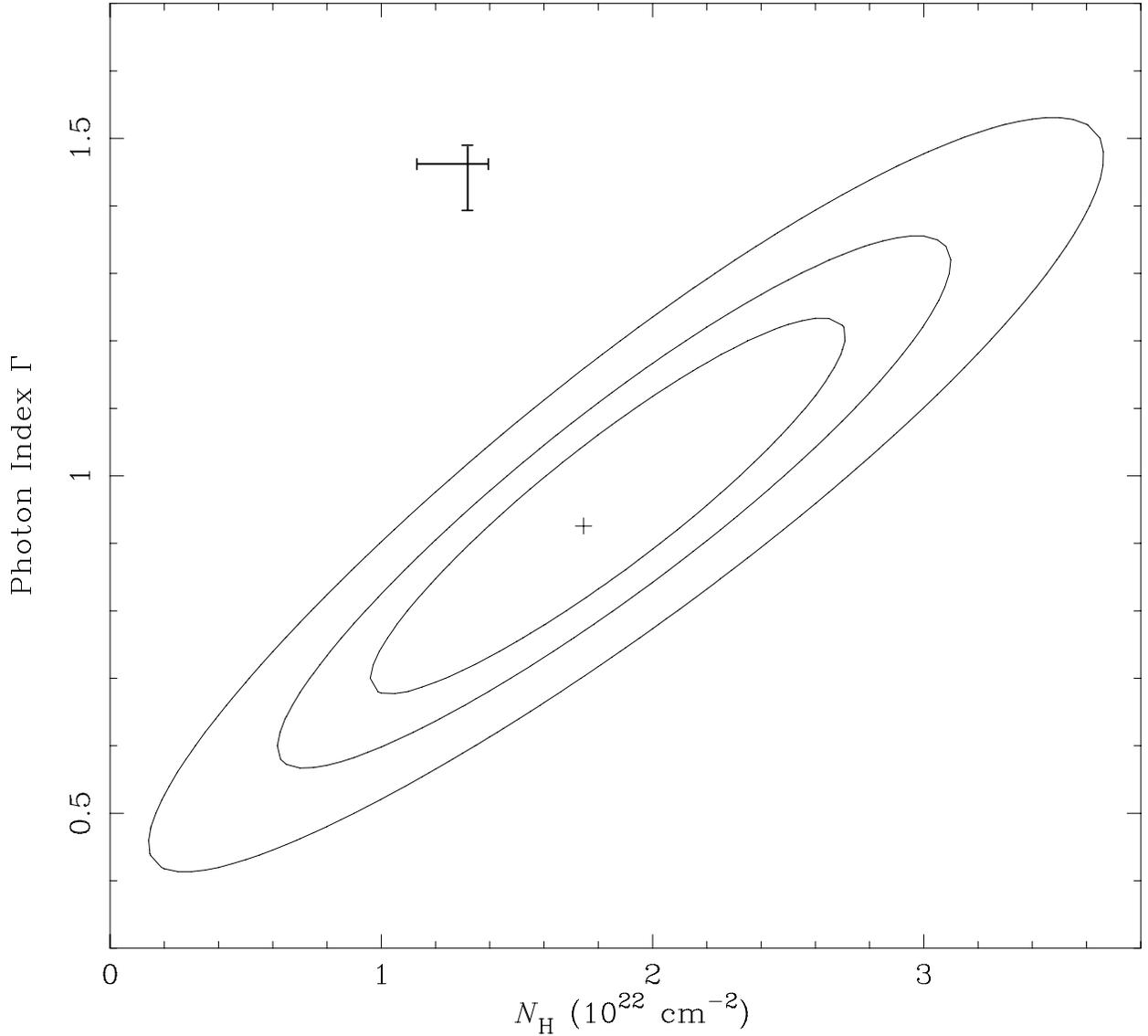}}
\caption{Confidence contours (68, 90, and 99\%) for a fit to the 2--9~keV
{\it Chandra\/} spectrum with a model consisting of a power law, Compton
reflection, and a relativistically broadened iron line.  The parameters
associated with the reflection and line components were fixed at the values
derived from {\sl ASCA\/} data by Shih et al.\ (2003); only the neutral
absorption $N_{\rm H}$ and power-law photon index $\Gamma$ were permitted
to vary in our fit.  The error bars indicate the range of values of
$N_{\rm H}$ and $\Gamma$ obtained by Shih et al.  The column density they
measured is consistent with our estimate, but the steep photon index they
obtained is clearly excluded in our fit, suggesting that the spectral slope
of the source varied between the {\it Chandra\/} and {\it ASCA\/}
observations.}
\end{center}
\end{figure}

\clearpage
\begin{figure}
\begin{center}
\centerline{\psfig{figure=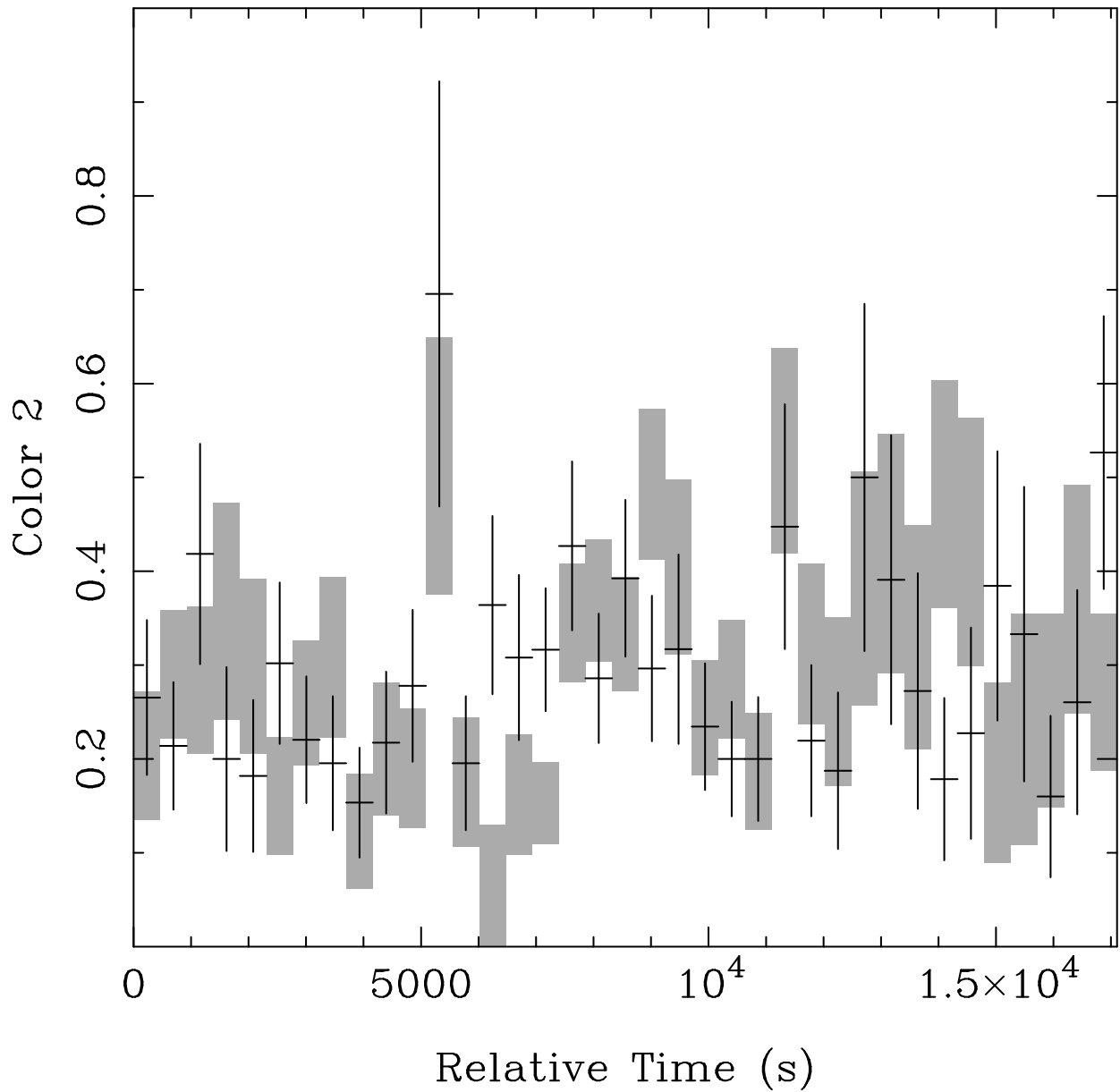,width=6.5truein,angle=0}}
\caption{Measured ({\it black crosses}) and predicted ({\it gray bars})
values of the Color~2 parameter.  The latter were derived from the observed
values of Color~1 shown in Fig.~3, under the assumption that color changes
arise from column density variations in the ionized medium alone.  The length
of the gray bars reflects the uncertainty in Color~1.}
\end{center}
\end{figure}

\end{document}